\documentclass{article}

\usepackage{arxiv}

\usepackage[utf8]{inputenc} 
\usepackage[T1]{fontenc}    
\usepackage{hyperref}       
\usepackage{url}            
\usepackage{booktabs}       
\usepackage{amsfonts}       
\usepackage{nicefrac}       
\usepackage{microtype}      
\usepackage{graphicx}
\usepackage{natbib}
\usepackage{doi}
\usepackage[doublespacing]{setspace}

\title{Molecular dynamics characterization of the interfacial structure and forces of the methane-ethane sII gas hydrate interface}

\date{} 					

\author{
    Samuel Mathews, Andr\'e Guerra, Phillip Servio, Alejandro Rey \\
	Department of Chemical Engineering
    McGill University \\
	Montreal, Quebec, Canada \\
}

\renewcommand{\headeright}{preprint}
\renewcommand{\undertitle}{
    This work has been submitted to be peer reviewed.
    Once the review process is complete, the DOI of the revised and published version will be added.
    This author's version is shared through a CC BY-NC-ND license.
}
\renewcommand{\shorttitle}{}

\begin{document}
\maketitle

\begin{abstract}
The nucleation of gas hydrates is of great interest in flow assurance, global energy demand, and carbon capture and storage.
A complex molecular understanding is critical to control hydrate nucleation and growth in the context of potential applications.
Molecular dynamics is employed in this work combined with the mechanical definition of surface tension to assess the surface stresses that control some of the behavior at the interface.
Ensuring careful sampling and simulation behavior, this work extracts meaningful results from molecular properties.
We characterize the interfacial tension for sII methane/ethane hydrate and gas mixtures for different temperatures and pressures.
We find that the surface tension trends positively with temperature in a balance of water-solid and water-gas interactions.
The molecular dipole shows the complexities of water molecule behavior in small, compressed pre-melting layer that emerges as a quasi-liquid.
These behaviors contribute to the developing knowledge base surrounding practical applications of this interface.
\end{abstract}


\newpage

\section{Introduction}
\label{sec:introduction}
Gas hydrates are clathrates where a backbone composed of water encloses gaseous molecules.
The water structure retains its integrity through hydrogen bonding and guest-host interactions between the gas and water\citep{carrollNaturalGasHydrates2014}.
Formed at low temperatures and high pressures, gas hydrates are commonly found in petroleum extraction and transportation industries\citep{carrollNaturalGasHydrates2014}.
These structures are presently studied in the context of methane extraction, desalination, flue gas capture, natural gas and hydrogen containment, carbon capture and storage technologies, and planetary ices\citep{kohFundamentalsApplicationsGas2011}.
With these technologies in mind, the need for critical information on the formation and dissociation of gas hydrates is highlighted.

The temperature, pressure, and guest identity are the main considerations governing hydrate formation, nucleation rate, growth rate, formation conditions, and morphology\citep{kohFundamentalsApplicationsGas2011}.
The three main structures are composed of combinations of large and small cages, with all three sharing a common pentagonal dodecahedron ($5^{12}$) cage.
The sI unit cell is cubic and composed of $5^{12}$ and $5^{12}6^2$ cages, and is formed by small gaseous molecules such as carbon dioxide, methane, and ethane.
The sII unit cell is also cubic and composed of $5^{12}$ and $5^{12}6^4$ cages, and is formed by propane, iso-butane, and natural gas mixtures.
The sH unit cell is hexagonal and composed of $5^{12}$, $4^35^66^3$ cages, and is formed by different combinations of large and small guests, such as methane and neohexene. 
Because natural gas is composed mainly of methane and ethane, the most common gas hydrate formed in flow assurance situations is the sII structure\citep{carrollNaturalGasHydrates2014}.
Considering the sII structure carefully is critical because at lower temperatures, only a small amount of ethane is required to favor it over the sI structure - at 273.15 Kelvin, some predictions show only 0.3 mole percent of ethane is required to stabilize sII.
Therefore, many hydrates that are believed to be sI may actually be sII hydrates.

There still exist many questions and unknowns on how gas hydrates form, what substances promote or inhibit their formation, what additives are best suited to induce their dissociation, and what exactly is happening at a molecular level during these processes.
Hydrates form and grow through various pathways, including as multiple crystals in agitated systems, as single crystals, and as films.
Typically, the nucleation starts at the gas/water interfaces and continue along the surface at a molecular scale, proving difficult to observe experimentally \citep{carrollNaturalGasHydrates2014}.
As the growth starts and continues, various behaviors are seen, including dendritic growth, shell formation, and other nanoscale structural dynamics\citep{mirzaeifardMultiscaleModelingSimulation2019, mirzaeifardCharacterizationNucleationMethane2019}.
These dictate the ideal location and pathway for gas hydrate formation, highlighting the need for accurate characterization of the molecular behaviors.

The molecular behaviors at the interface are key features in area that are concerned with nucleation, phase transitions, solubilization of fluids, the design of surface-controlled materials such as soaps and adhesives, and the study complex hydrogen bonding systems in cell biology\citep{carrollNaturalGasHydrates2014}.
The main drivers of the formation mechanisms are the driving forces and interfacial energy, but thorough atomic characterization of the processes remains incomplete\citep{blasVaporliquidInterfacialProperties2008}.
The speed and size of the dynamics of systems at this scale give rise to challenges that can restrict experimental methods. 
For example, in the study of protein hydration, traditional structural description methods are not suitable at the interface due to breaking of the translational and rotational symmetry preservation\citep{shiStructuralOrderProtein2022}.

Therefore, the use of computational materials science techniques such as molecular dynamics (MD) have given researchers opportunities to glimpse into the nucleation process and understand the molecular behaviors\citep{mirzaeifardMolecularDynamicsCharacterization2019}. 
The motivation of this computational work is to qualify the interfacial tension of methane-ethane sII gas hydrates in the presence of natural gas and reveal the important structures and physics in these systems.
This type of vapor/solid system has been studied for other structures in the context of characterizing nanobubbles and interfacial gas enrichment.
The behavior of this layer is critical to the bubble's stability and is responsible for many phenomena in atmospheric chemistry, the motion of glaciers, and even the microstructure of snow\citep{kohFundamentalsApplicationsGas2011,guerraIntegratedExperimentalComputational2022}.

The organization of this paper is as follows.
In the methodology, the general computational details will be presented, followed by a description of the MD and numerical techniques used to calculate the various systems and surface properties.
Then, the results are presented, including experimental and theoretical data for validation and discussion of the results.
We calculate and analyze the surface tension of the hydrate-gas system, characterize the interfacial structure and ordering by analyzing the molecular dipole and the interfacial thickness, and scrutinize the organization of the system by considering the density distributions.

\section{Methodolody}
\label{sec:methodology}

\subsection{General Computational Details}
\label{sec:methodology:general}
We employ MD as implemented in the Large-scale Atomic/Molecular Massively Parallel Simulator (LAMMPS)\cite{thompsonLAMMPSFlexibleSimulation2022}.
It has been successfully used in the past to study a variety of hydrate interfacial systems\cite{mirzaeifardCharacterizationNucleationMethane2019, mirzaeifardMolecularDynamicsCharacterization2018, mirzaeifardMolecularDynamicsCharacterization2019, naeijiInterfacialPropertiesHydrocarbon2019, naeijiMolecularDynamicsSimulations2020}.
The system is composed of solid sII methane/ethane hydrate, where small cages are filled with methane and large cages are filled with ethane, surrounded by 95\% methane and 5\% ethane gas.
The hydrate structure is 100\% occupied.
This configuration improves property sampling by allowing the average value of the given property to be calculated from two interfaces.
The system is simulated from 15.0 to 40.0 atmospheres and 273.15 to 298.15 Kelvin. 
This narrow band of conditions is necessary because at low temperatures, small amount of ethane can destabilize the sII hydrates.
As a consequence, some permafrost and deep sea deposit thought to be composed of sI hydrates may actually be sII hydrates\cite{ballardOptimizingThermodynamicParameters2000}.

\begin{figure}
    \centering
    \includegraphics[width=\linewidth]{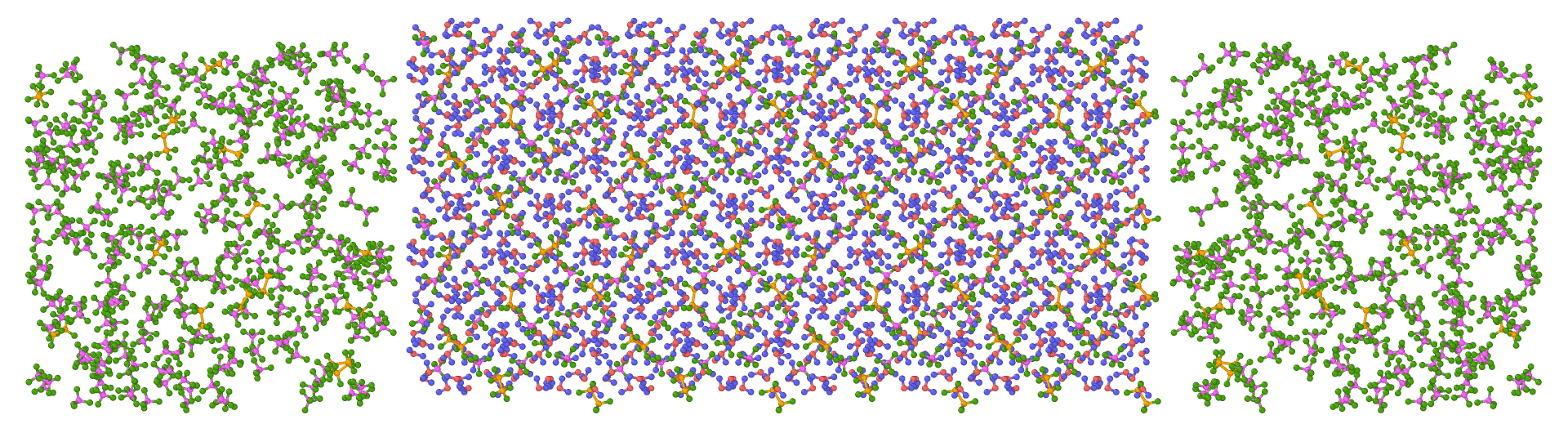}
    \caption{Initial configuration of the simulation box. Red atoms represent oxygen, blue atoms represent hydrogen bonded to the oxygen, green atoms represent hydrogen in hydrocarbons, purple atoms represent carbon in methane, orange atoms represent carbon in ethane. The gap between phases exists to improve simulation behavior during the initial temperature shock.}
    \label{fig:system_initial}
\end{figure}

Initially, the simulation box has the dimensions of $40\times140\times40$ \AA\ in the $x$, $y$, and $z$ directions, corresponding to a solid phase of 16 units cells of the sII structure, each with a lattice parameter of 17.31 \AA.
The clathrate's atomic coordinates are obtained from X-ray diffraction to satisfy the Bernal-Fowler ice rules, providing the lowest potential energy configuration\cite{takeuchiWaterProtonConfigurations2013}.
The gas boxes each contain 285 methane molecules and 15 ethane molecules to recreate realistic natural gas composition.
The simulation box contains 2176 water molecules, 826 methane molecules, and 158 ethane molecules.
Figure \ref{fig:system_initial} shows the initial configuration of the system. 

Gas molecules in the bulk phase are placed randomly, and the guests in the hydrate phase are randomly oriented, but centered in each cage.
The random orientation and placement creates appropriate initial configurations for MD by minimizing short range repulsive interactions with the use of PACKMOL to pack the molecules\cite{martinezPACKMOLPackageBuilding2009} with a separation tolerance of 2.0 \AA and reduces computational cost.

Moltemplate was used to assign force field parameters and associated values\cite{jewettMoltemplateToolCoarseGrained2021}.
The Optimized Potentials for Liquid Simulations All Atom (OPLSAA) force field provided the best choice for the hydrocarbons\cite{jorgensenMolecularModelingOrganic2005}.
For the water molecules, the TIP4P/Ice force field provided the best performance in the ice phase and was chosen for the hydrate structure elements\cite{abascalPotentialModelStudy2005}.
Despite being optimized for water in the solid phase, this potential has been shown to perform well in the vicinity of interfaces and for liquid water.
The relevant force field parameters are found in Table \ref{tab:ff_params}.
A cutoff of 12 \AA\ was used for both the Lennard-Jones (LJ) and Coulombic interactions, with the arithmetic Lorentz-Berthelot mixing rule applied for inter-molecular interactions:
\begin{equation}
    \label{eqn:lj}
    E=4\epsilon\left[\left(\frac{\sigma}{r}\right)^{12}-\left(\frac{\sigma}{r}\right)^{6}\right]\quad r<r_c
\end{equation}
\begin{equation}
    \label{eqn:coul}
    E=\frac{Cq_{i}q_{j}}{\epsilon{r}}\quad r<r_c
\end{equation}
\begin{equation}
    \label{eqn:epsilon}
    \epsilon_{ij}=\sqrt{\epsilon_{ii}\epsilon_{jj}}
\end{equation}
\begin{equation}
    \label{eqn:sigma}
    \sigma_{ij}=\frac{\sigma_{ii}+\sigma_{jj}}{2}
\end{equation}

Angles were treated with the harmonic style.
The angles and bond length of the water molecules were constrained using the Shake algorithm.
The long-range particle-particle particle-mesh solver of Hockney and Eastwood\cite{hockneyComputerSimulationUsing1988} was employed with a force error accuracy of $1 \times 10^{-5}$.
Some post processing steps were performed using MDAnalysis, a Python library useful for parsing, handling, and processing trajectories output from LAMMPS\cite{michaud-agrawalMDAnalysisToolkitAnalysis2011, gowersMDAnalysisPythonPackage2016}.
We integrate the non-Hamiltonian equations of motions with the Verlet algorithm with a 2 femtosecond time step.
The Nos\'e-Hoover barostat and thermostat is employed to control the pressure and pressure.
The simulations ran for 200 nanoseconds under the NPT ensemble.
Then, the system was run under the NVT ensemble to ensure structural stability without the barostat.
The quantities of interest were averaged over the last 5 nanoseconds of simulation time.

\begin{table}
    \label{tab:ff_params}
    \caption{Simulation parameters and properties of simulation molecules}
    \centering
    \begin{tabular}{llrl}
        & \textbf{Quantity} & \textbf{Value} & \textbf{Units} \\
        \hline 
        \textbf{H$_2$O} & O mass & 15.999 & g${\cdot}$mol$^{-1}$ \\
        & O charge & -1.1794 & e \\
        & H mass & 1.008 & g${\cdot}$mol$^{-1}$ \\
        & H charge & 0.5897 & e \\
        & OO $\epsilon$ & 0.21084 & kcal$\cdot$mol$^{-1}$ \\
        & OH, HH $\epsilon$ & 0.0 & kcal$\cdot$mol$^{-1}$ \\
        & OO $\sigma$ & 3.1668 & \AA \\
        & OH, HH $\sigma$ & 0.0 & \AA \\
        & OH r$_0$ & 0.9572 & \AA \\
        & HOH $\theta{_0}$ & 104.52 & \textdegree \\
        \hline
        \textbf{CH$_4$/C$_2$H$_6$} & C mass & 12.011 & g${\cdot}$mol$^{-1}$ \\
        & H mass & 1.008 & g${\cdot}$mol$^{-1}$ \\
        & CC $\epsilon$ & 0.066 & kcal$\cdot$mol$^{-1}$ \\
        & HC $\epsilon$ & 0.03 & kcal$\cdot$mol$^{-1}$ \\
        & CC $\sigma$ & 3.5 & \AA \\
        & HC $\sigma$ & 2.5 & \AA \\
        \hline
    \end{tabular}
\end{table}

\subsection{Interfacial Properties}
\label{sec:methodology:interfacial}
According to fundamental thermodynamics, the surface free energy may be defined as the isothermal work of formation of a unit area of an interface.
According to molecular theory, the surface free energy can also be defined mechanically as the stress felt across a strip of unit width normal to the Gibbs dividing surface of an interface\cite{kirkwoodStatisticalMechanicalTheory1949}.
Therefore, the surface tension of an interface can be defined by the following, considering the $y$ direction to be normal to the interface in equation and the fact that the system has two interfaces:
\begin{equation}
    \label{eqn:gamma_kb}
    \gamma=\frac{1}{2}\int_{-\infty}^{+\infty}\left(P_N-P_T(y)\right)\,dy
\end{equation}
where $P_N$ is the normal pressure and $P_T$ is the tangential pressure.
The terms in equation \ref{eqn:gamma_kb} are related to the stress tensor obtained from MD simulations\cite{rowlinsonMolecularTheoryCapillarity2002}:
\begin{equation}
    \label{eqn:pt}
    P_T=\frac{1}{2}(P_{XX}+P_{ZZ})
\end{equation}
\begin{equation}
    \label{eqn:pn}
    P_N=P_{YY}
\end{equation}
where $P_{ii}$ is the $i$th element of the diagonal of the tensor.
This yields the following for surface tension:
\begin{equation}
    \label{eqn:gamma}
    \gamma=\frac{1}{2}\int_{-\infty}^{+\infty}\left(P_{YY}-\frac{1}{2}(P_{XX}+P_{ZZ})\right)\,dy
\end{equation}

Due to the finite nature of MD simulations, a correction is required to account for the cutoff using in the LJ potential.
Otherwise, the mechanical definition in equation \ref{eqn:gamma} underestimates the surface tension\cite{chapelaComputerSimulationGas1977, blokhuisTailCorrectionsSurface1995}.
The correction is made according to the formula of Chapela et al.\cite{chapelaComputerSimulationGas1977} and improved upon by Blokhius et al.\cite{blokhuisTailCorrectionsSurface1995}:
\begin{equation}
    \label{eqn:tail}
    \gamma_{tail}=\int_{0}^{1}\int_{r_c}^{\infty}12\pi\epsilon\sigma^6(\rho_h-\rho_g)^2\left(\frac{3s^3-s}{r^3}\right)coth\left(\frac{rs}{d}\right)\,dr\,ds
\end{equation}
where $r_c$ is the cut-off distance, $\rho_h$ is the molecular density of the hydrate phase, $\rho_g$ is the molecular density of the gas phase, and $d$ is the interfacial thickness.
The effect of including long range corrections during the simulation is small when interfacial properties are calculated, and applying the correction afterwards to the surface tension is a suitable way to make the correction\cite{sakamakiThermodynamicPropertiesMethane2011}.
Dealing with the truncation of potentials in this way in systems exhibiting planar symmetry has been shown to be effective\cite{martinez-ruizEffectDispersiveLongrange2014}.
The large LJ and Coulombic cutoff ensures a small correction.

With equations \ref{eqn:gamma} and \ref{eqn:tail}, the total tension is:
\begin{equation}
    \label{eqn:gammaT}
    \gamma_{total}=\gamma+\gamma_{tail}
\end{equation}
While the integration must be performed to evaluate equation \ref{eqn:tail}, for equation \ref{eqn:gamma_kb} one can use the average values for the pressure tensor elements for the simulation box.

\section{Results and Discussion}
\label{sec:results}

\subsection{Equilibration}
\label{sec:results:equilibration}
In MD, questions arise regarding sampling systems consisting of many measurable parameters of varying length and time scales.
Therefore, the first step in any MD study should be to assess if the system is at conditions favorable to adequate sampling.
In the case of this study, all systems should be at equilibrium.
Two strategies for assessing the behavior of the system nearing equilibrium are analyzing the time series of some key quantity to assess fluctuations and using distance measures over the trajectory of the MD system to evaluate the sampling of the configurational space\cite{grossfieldBestPracticesQuantification2019}.

For the first strategy, the time series of various scalar values is observed to check for drift.
Common values are pressure, temperature, system volume, temperature, and/or density\cite{guerraAllAtomMolecularDynamics2022, guerraMolecularDynamicsPredictions2023}.
In all cases, the simulations showed that the values fluctuate rapidly above and below their mean value, meaning the system is sampling different states corresponding to different values of the parameter space.
The lack of drift with simulation time indicates equlibrium.

\begin{figure}
    \centering
    \includegraphics[width=0.5\linewidth]{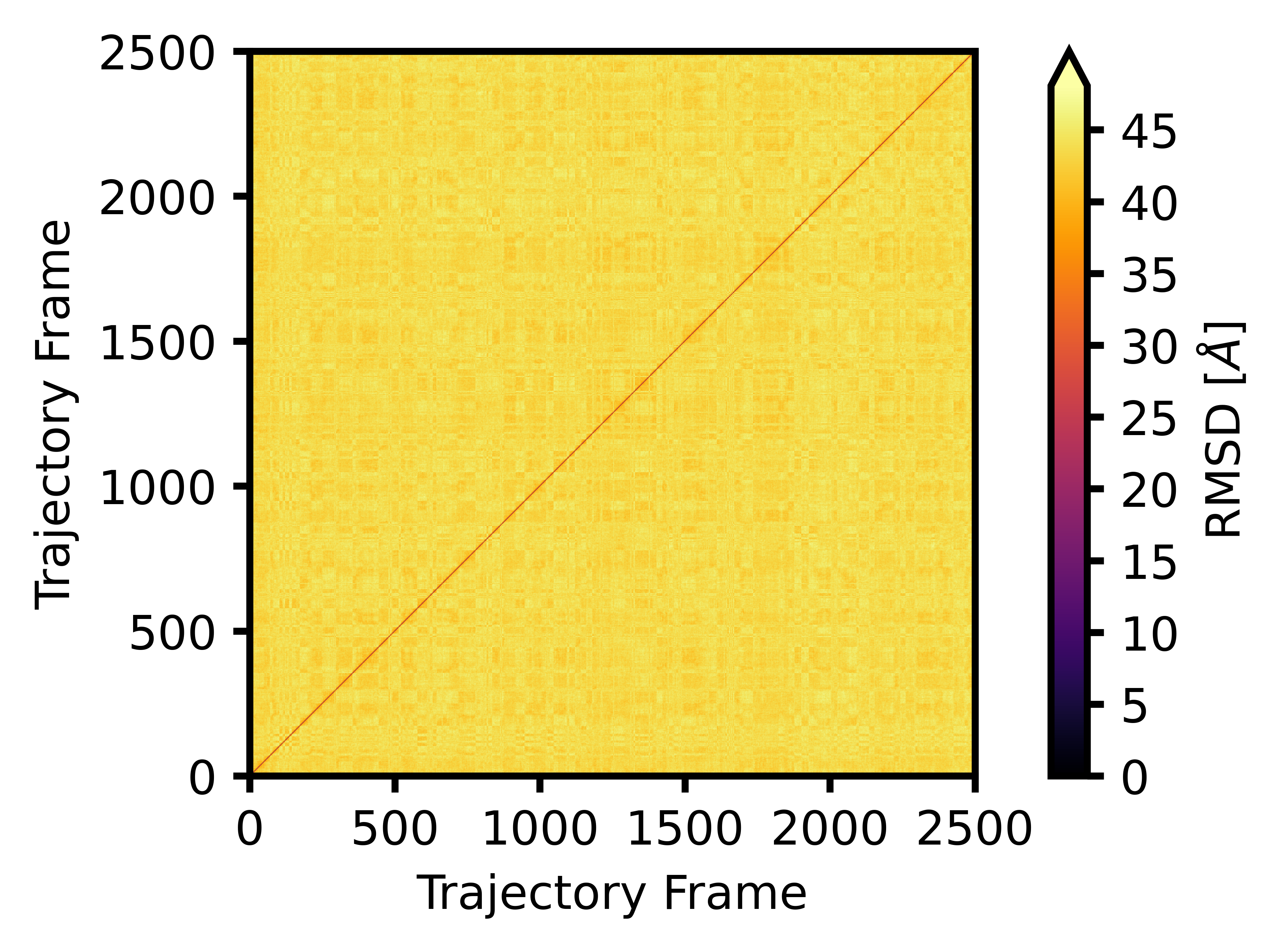}
    \caption{All-to-all root mean square deviation (RMSD) of the simulation trajectory. As the distance from every other frame is large, the simulation has sampled enough states to solidy the claim of equilibrium. These plots are for 30.0 atmospheres and 283.15 Kelvin. Snapshots are taken every 4 picoseconds.}
    \label{fig:rmsd}
\end{figure}

The second strategy uses distance based metrics to assess how similar molecular configurations are to each other.
The all-to-all root mean square deviation (RMSD) is the square root of the average of the square of the difference in position of all atoms during the simulation. 
By comparing all trajectory frames to each other (all-to-all), similarity between frames is assesed.
Figure \ref{fig:rmsd} shows one all-to-all RMSD computed.
The diagonal always shows 0 RMSD.
Darker regions indicate pockets of similar structures.
Off diagonal RMSD minimums show that the system is sampling configurational states that have already been visited, indicating proper sampling\cite{grossfieldBestPracticesQuantification2019}.
Therefore, we can conclude that our system is at equilibrium in the context of molecular simulation.

\subsection{Interfacial Structure}
\label{sec:results:interfacial}
\begin{figure}
    \centering
    \includegraphics[width=\linewidth]{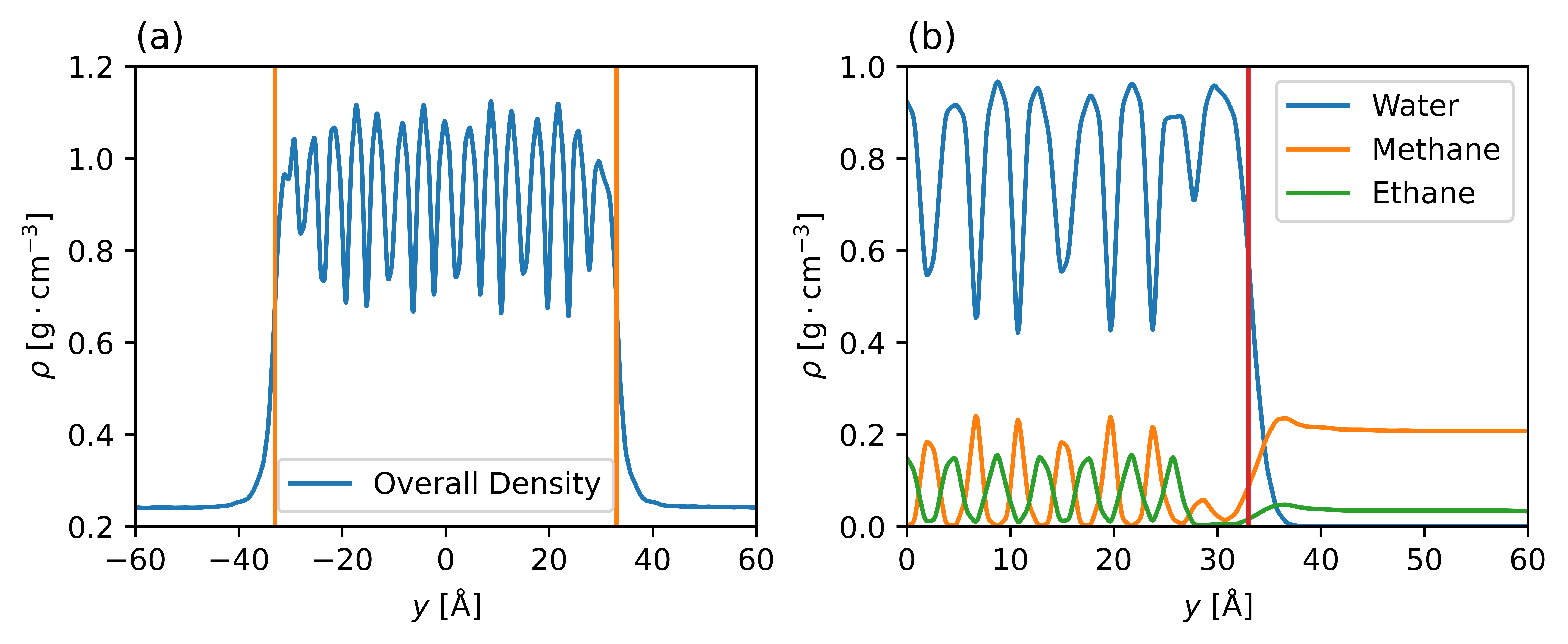}
    \caption{(a) shows the density of the system along the axis normal to the interface. (b) shows the rightmost interface of the same system and the axial density distribution of the three components. Both show peaks and valleys corresponding to the periodic nature of the system. The vertical lines represent the center of the interface as defined by the hyperbolic tangent regression. These plots are for 30.0 atmospheres and 283.15 Kelvin. The density peaks for the methane and ethane in (b) demonstrate gas accumulation at the edge of the pre-melting layer.}
    \label{fig:rho}
\end{figure}

The first step in calculating the interfacial tension is to compute the density profile to perform the tail correction.
The density profile also provides insight into the interfacial structure.
The axial density profile of the system is shown in figure \ref{fig:rho}, showing oscillations in the hydrate phase, expected in periodic solids.
Plot (b) focuses on one interface. 
Starting from the center of the hydrate phase and moving towards the gas, there is a drastic decline in the overall density towards the bulk value.
While steep, this decline still occurs over some spatial distance and is indicative of a pre-melting, connective interfacial layer that relates the properties of the bulk gaseous phase to the solid hydrate phase.

Characterizing the pre-melting layer is complex due to its size and nature.
Plot (b) of figure \ref{fig:rho} shows that the interfacial layer is composed of semi-liquid water as there is a transition from cage to liquid to gas.
The gas molecules either are in the hydrate or diffuse to the bulk, with near zero density in the layer. 
Additionally, there is adsorption of methane and ethane molecules on the hydrate surface due to the heterogeneous nucleation shown by the peaks in methane and ethane densities that exceed the bulk densities.
This layer and its behavior has been seen in similar system types before\cite{mirzaeifardCharacterizationNucleationMethane2019, mirzaeifardMultiscaleModelingSimulation2019}.
It is critical in fields such as atmospheric chemistry, the dynamics of glacier motion, and even the nanostructure of snow\cite{qiuWhyItDifficult2018}.

To use the tail correction, we require the bulk densities and physical thickness of the interface.
We use a hyperbolic tangent function to determine the bulk densities and physical thickness of the interface, which are critical for the tail correction\cite{blasVaporliquidInterfacialProperties2008}.
\begin{equation}
    \label{eqn:tanh}
    \rho_{ij}(y)=\frac{1}{2}(\rho_i+\rho_j)-\frac{1}{2}(\rho_i-\rho_j)\mathrm{tanh}\left(\frac{|y-y_c|-y_G}{d}\right)
\end{equation}
In equation \ref{eqn:tanh}, $y$ is along the axis normal to the interface, $d$ is the physical thickness, $y_c$ is halfway between the two interfaces, $y_G$ is the Gibbs dividing surface position, and $\rho_i$ and $\rho_j$ are the bulk densities of the two different phases in question.
The densities are in units of $\mathrm{g\cdot cm^{-3}}$ and the distances are in units of \AA.
$y_c$, $y_G$, $d$, and the two bulk densities are the fitting parameters.
The two bulk densities in equation \ref{eqn:tanh} match the bulk densities directly measured in the system, confirming the findings of the regression procedure.
$y_c$ is the center of the hydrate phase, and helps normalize the spatial coordinates to make them relative to the phase and not to the simulation box, attenuating small fluctuations in the system positions.

\begin{figure}
    \centering
    \includegraphics[width=\linewidth]{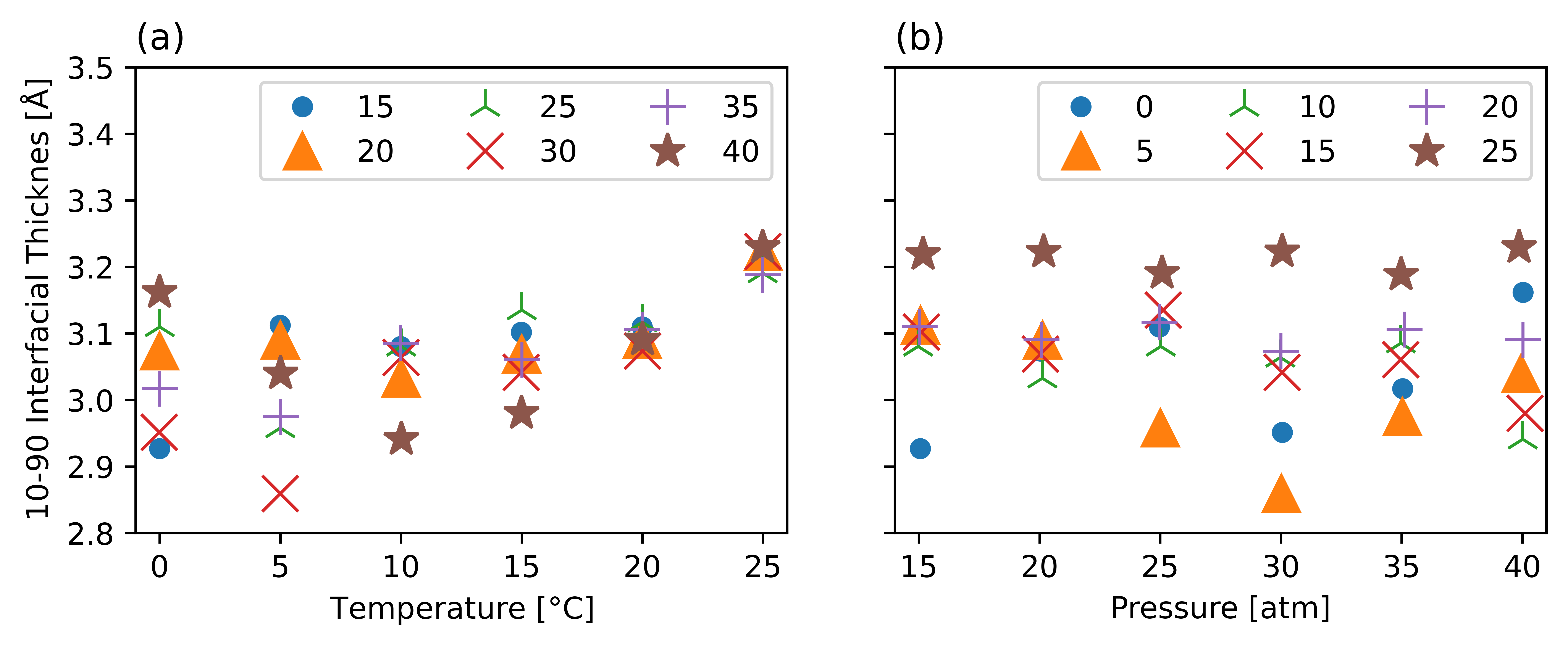}
    \caption{(a) shows the 10-90 interfacial thickness of the system as a function of temperature for a given set of pressures in atmospheres. (b) shows the 10-90 interfacial thickness of the system as a function of pressure for a given set of temperatures in Celsius. (a) shows the tendency of the interface to expand with temperature and (b) shows the interfaces is comparatively insensitive to pressure over the range.}
    \label{fig:thickness}
\end{figure}

Another measure of the interfacial thickness is the 10/90 thickness, which is related to the regressed physical thickness $d$ by $t=2.1972d$\cite{blasVaporliquidInterfacialProperties2008}.
The 10/90 thickness $t$ is less sensitive to the shape of the density profile than $d$, but it cannot be estimated with appreciable precision from results in computer simulations that may have noise or specific periodicity as in our case\cite{rowlinsonMolecularTheoryCapillarity2002}.
The physical thickness $t$, estimated using the regressed physical thickness $d$, for all conditions with respect to pressure and temperature is presented in figure \ref{fig:thickness}. 
The trend with pressure is not pronounced for stable sII hydrates.
For sII natural gas hydrate systems, the interfacial thickness increases with temperature, consistent with the expansion of the simulation box as temperature increases combined with the relatively low thermal expansion of hydrate crystals\cite{mathewsHeatCapacityThermal2020} leading to the spatial differences coming from the expansion of the gaseous and quasi-liquid layer.
Results from this work show that the interfacial thickness as regressed or the 10/90 thickness in natural gas and hydrate interfaces are smaller than for water-methane interfaces\cite{mirzaeifardMolecularDynamicsCharacterization2018} and for water-ethane and water-propane interfaces\cite{mirzaeifardMolecularDynamicsCharacterization2019}.

Quantifying the sensitivity of the regressed thickness to the averaging length showed that there is a strong correlation between the thickness and the length when it is larger than a few \AA\ .
Without this spatial averaging, the interfacial thickness cannot be regressed with physical meaning from point density measurements.
By using 1 \AA\ as our chunk size, we could support mesh independence, which has not been explicitly stated in much literature surrounding these measurements and regressions in hydrates.
These issues create a need for a better way to characterize the thickness of the interface.

The axial charge distribution provides one method to characterize the interfacial region, but it lacks directional information that may be helpful in determining some more nuanced distributions in such a small and chaotic region.
It has been used in the past\cite{mirzaeifardCharacterizationNucleationMethane2019, mirzaeifardMolecularDynamicsCharacterization2018, mirzaeifardMolecularDynamicsCharacterization2019, naeijiInterfacialPropertiesHydrocarbon2019, naeijiMolecularDynamicsSimulations2020} but it still suffers from the same shortcomings of the density fitting procedures.
The organization of molecules at solid surfaces is critical in material behaviors like vibrational dynamics, surface conductivity, chemical reaction, and adsorption.
The liquid pre-melting layer at the interface's organization depends on the balance between water-water, water-gas, and water-hydrate interactions.
The molecular dipole can help quantify these interactions.

\begin{figure}
    \centering
    \includegraphics[width=0.5\linewidth]{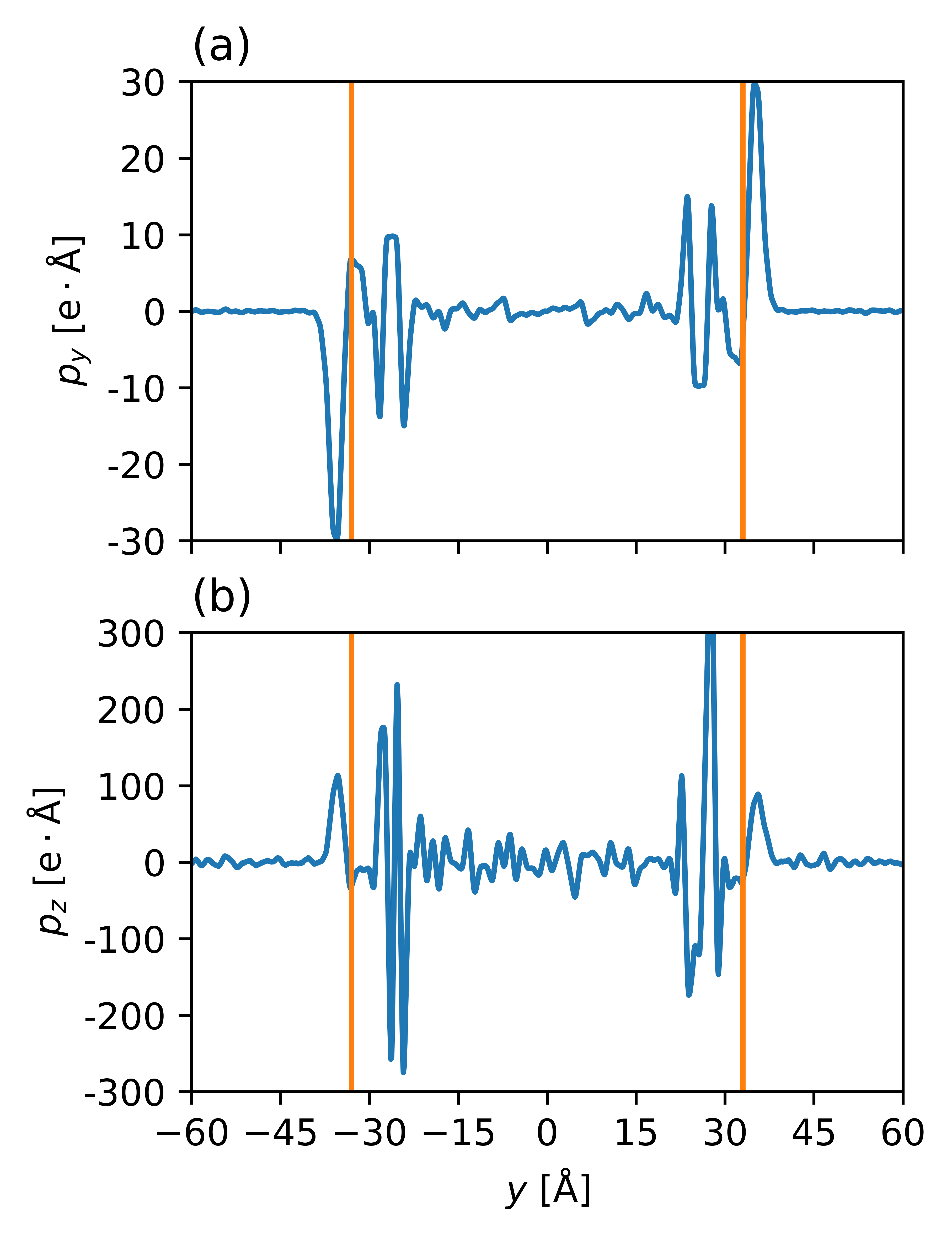}
    \caption{(a) shows the y-component (normal to the interface) of the dipole vector as a function of position along the axis normal to the interface. (b) shows the z-component (tangent to the interface) of the dipole vector as a function of position along the axis normal to the interface. These plots are for 30.0 atmospheres and 283.15 Kelvin. The orange line represents the center of the interface. (a) shows the tendency of water molecules to align themselves with the oxygen molecules pointing towards the solid phase. (b) shows the strong tendency of water molecules to be aligned parallel to the solid phase as well. Intense peaks at $|y|<33$ \AA\ indicate strong alignment in layers.}
    \label{fig:dipole}
\end{figure}

Keeping it mind that the hydrocarbons herein do not possess a net dipole and that the dipole is defined as thepointing from negative to positive, we study the spatial dependence of dipole componenets.
Figure \ref{fig:dipole} shows the y-component (normal to the interface) and the z-component (tangent to the interface) of the system as a function of the y-axis throughout the system. 
Plot (a) shows that the water molecules align themselves with the oxygen atoms facing toward the hydrate structure very close to the interface. 
However, plot (b) shows that there is also a strong alignment parallel to the interface.
Within the pre-melting layer, there are some more chaotic behaviors as the water molecules start to participate in the organize clathrate structure.

Figure \ref{fig:dipole} shows that there are strong organizational behaviors at play right at the interface and in the pre-melting layer, but that they dissipate quickly in the bulk and within a few molecular layers in the interface, as has been seen in water/hydrocarbon mixtures analyzing the dipole at the interface\cite{fernandesMolecularDynamicsSimulation1999}. 
In particular, plot (b) indicates that the water molecules are aligned in layers prior to their participation in the hydrate structure to provide the highest probability that they can hydrogen bond with the clathrate backbone as the hydrogen or oxygen atoms could participate in a hydrogen bond.
The weak correlation of interfacial thickness and density profiles with temperature agrees with estimates based on molecular dipoles and shows that in fact this gas/solid interface is narrow.

\begin{figure}
    \centering
    \includegraphics[width=0.5\linewidth]{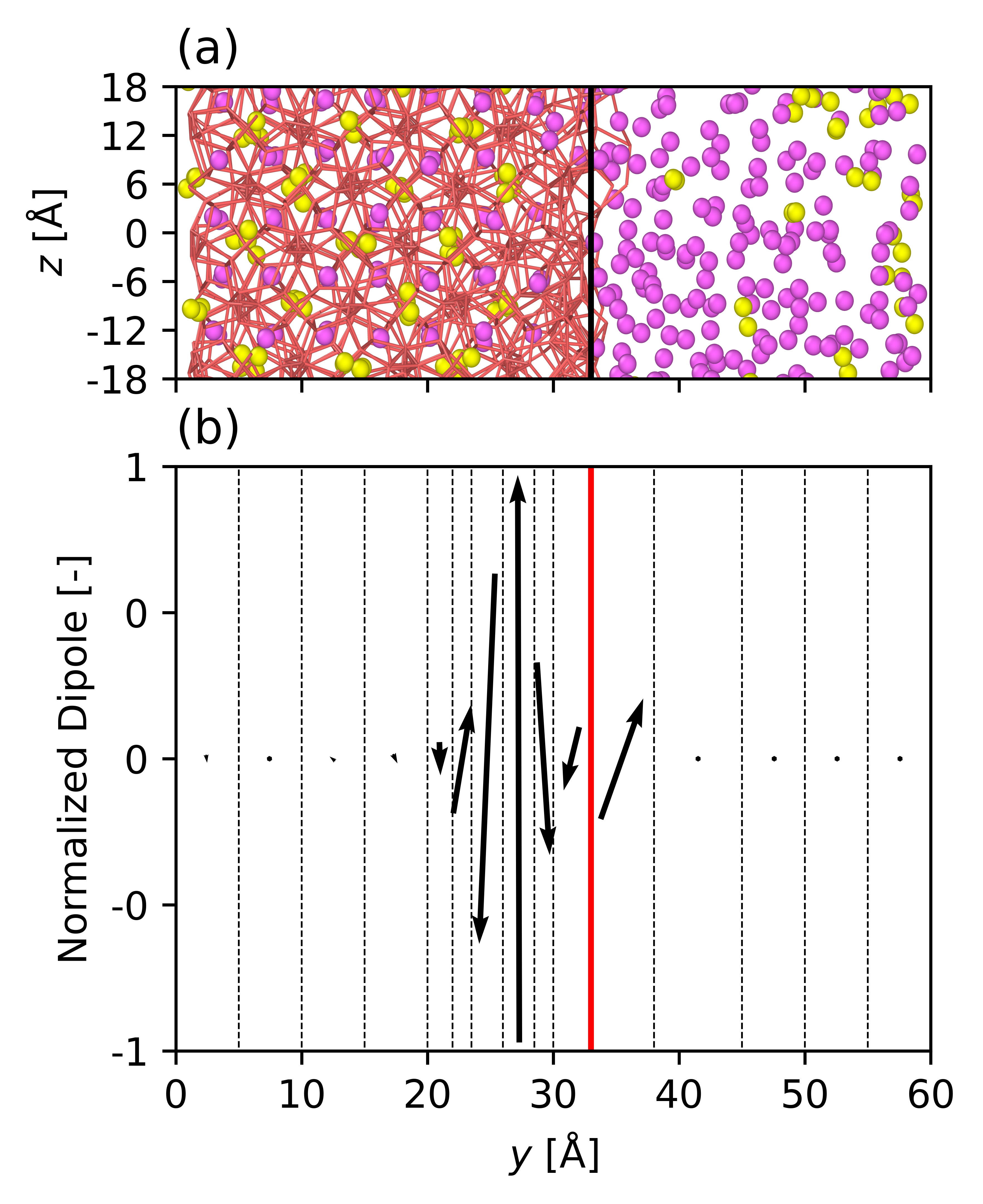}
    \caption{(a) shows a snapshot of the system configuration at the final timestep. The water molecules are suppressed and the red links are between oxygen atoms to show the cage structure of the hydrate. The yellow and pink spheres represent ethane and methane molecules, respectively. The black line shows the center of the interface. (b) shows the normalized net dipole vector normalized for each spatial region between two dashed lines. The red line represents the center of the interface. These plots are for 30.0 atmospheres and 283.15 Kelvin. (b) constitutes an average of the two interfaces present in the system.}
    \label{fig:combo}
\end{figure}

Figure \ref{fig:combo} shows the dipole moments of different chunks along the axis normal to the interface but only for one side of the simulation box.
The magnitude of the dipole component that corresponds to the parallel alignment is very strong within the pre-melting layer closest to the solid phase, but the water molecules in the layer furthest from the center (at $y=33.4$ \AA) have some component normal to the interface due to the competing surface charges in the pre-melting layer.
The water oxygen atoms prefer the charge corresponding to the water confined in the layer that is not participating in the hydrate structure, whereas the hydrogen atoms prefer the opposite dipole arrangement with the bulk phase.
Such behaviors exist in carbon nanotubes and confined water systems, where the positive surface charge of the tube requires the first layer of water to be aligned with the oxygen atoms towards the solid\cite{zhaoRoleInterfaceIons2019}.

Figure \ref{fig:combo} also displays the strength of the parallel alignment once the water molecules have moved into the pre-melting layer. 
The tangent component nearly completely dominating the directional tendencies of the water molecule.
It is strongest in the region of the clathrate that is beginning to melt, with the methane and ethane densities being at the lowest points where the water dipole is dominantly in its parallel alignment.
These behaviors were seen for all conditions examined, across all pressures and temperatures.
Biscay et al.\cite{biscayAdsorptionNalkaneVapours2011} found that hydrocarbon display parallel orientations to water surfaces in hydrocarbon/water systems, as did Naeiji et al. for hydrocarbons\cite{naeijiInterfacialPropertiesHydrocarbon2019} and for carbon dioxide molecules\cite{naeijiMolecularDynamicsSimulations2020}.
However, the dipole behavior of water molecules in hydrate systems and their precursors remains unexamined until now.
Studying the dipole is critical in many engineering phenomena, as it can indicate adsorption processes, dielectric responses, surface conductivity, and mass transfer.
The balance of the forces that is clarified by studying the dipole sheds light on water-solid and water-water interactions, explaining the hydrophobicity and hydrophilicity of surfaces at the atomic scale.

\subsection{Interfacial Tension}
\label{sec:results:tension}

To quantify the forces and stresses of gas hydrate systems, we calculate the interfacial tension for the contact surface. 
It is critical to note that the scope of this research project is centered around understanding the forces, tensions, and stresses at the interface, and does not extend to the energy of the interface.
For surfaces not involving solids, the surface free energy and surface tension are numerically equivalent.
However, for solids, there is a difference that can be captured by the Shuttleworth equation\cite{shuttleworthSurfaceTensionSolids1950}.
Detailed discussion on the differences between the two calculations and how to untangle common misconceptions has been provided by Di Pasquale et al.\cite{dipasqualeShuttleworthEquationMolecular2020} and Hecquet\cite{hecquetSurfaceEnergySurface2018}.

\begin{figure}
    \centering
    \includegraphics[width=\linewidth]{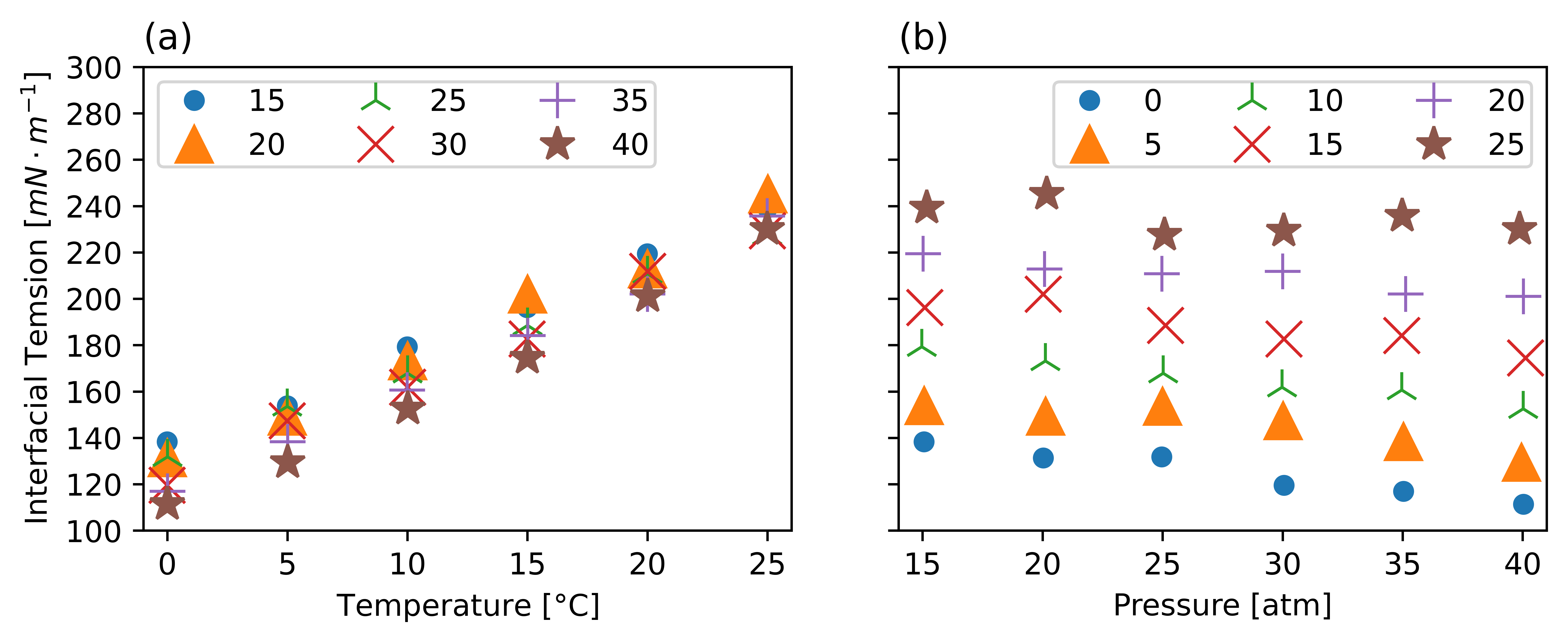}
    \caption{(a) shows the interfacial tension of the system as a function of temperature for a given set of pressures. (b) shows the interfacial tension of the system as a function of pressure for a given set of temperatures. The strong correlation with temperature in (a) is similar to water/hydrate liquid/solid systems. The weak correlation in (b) is common is other hydrate and hydrate precursor systems\cite{mirzaeifardMultiscaleModelingSimulation2019, mirzaeifardMolecularDynamicsCharacterization2019, mirzaeifardMolecularDynamicsCharacterization2018}.}
    \label{fig:tension}
\end{figure}

Figure \ref{fig:tension} reports the surface tension for the natural gas hydrate system in contact with natural gas. 
There is scarce existing work in the relevant field, but some measurements for sI hydrate/gas mixtures have been simulated by Mirzaeifard et al.\cite{mirzaeifardCharacterizationNucleationMethane2019}, finding that the interfacial tension is largely insensitive to pressure and temperature for these systems, hovering at 95 $mN\cdot m^{-1}$ for sI hydrate and methane mixtures.
Figure \ref{fig:tension} shows that while this is true for pressure, for temperature there is a positive correlation.

Correlations of surface tension with temperature for hydrocarbon/water, hydrate/water, and hydrate/gas systems are complex and are not in agreement across literature.
While some findings follow the general trend that surface tension decreases with increasing temperature for hydrocarbon/water systems\cite{naeijiInterfacialPropertiesHydrocarbon2019}, others find results of increasing surface tension with temperature in methane/water mixtures and carbon dioxide/water mixtures\cite{naeijiMolecularDynamicsSimulations2020}.
Experimental results for ice-water-air line tension show an increase with temperature while the ice-air interfacial tension shows the same characteristics\cite{djikaevSelfConsistentDeterminationIce2017}.

Mirzaeifard et al.\cite{mirzaeifardMultiscaleModelingSimulation2019} found that while the interfacial energy decreases with temperature, the interfacial tension, defined as the sum of the interfacial energy and the product of the interfacial area and the derivative of energy with respect to area, increases with temperature for water/methane hydrate systems, ranging from 30 $mN\cdot m^{-1}$ at 271 Kelvin to 35 $mN\cdot m^{-1}$ at 290 Kelvin. 
These results were calculated with the OPLS-UA force fields which consider methane molecules as single particles.
Additionally, the TIP4P force field for water was employed, not the TIP4P/Ice force field, and therefore the properties differ thanks to the optimization differences between the two.
This corresponds to results for natural gas/hydrate systems in this study, and follows the definitions of the surface tension or stress for solids when compared to the surface energy.
When considering the surface entropy from fundamental thermodnamics as the negative of the partial derivative of surface tension with respect to temperature, assuming constant pressure and area, we arrive at the conclusion that, as the temperature increases, the disorder is decreasing. 
As the temperature goes up, the surface entropy is weakened and the well-ordered hydrate phase is inhibiting the free rotation of molecules, particularly water.
Such behavior does not follow classical entropic behavior seen for liquid phases\cite{mirzaeifardMolecularDynamicsCharacterization2018}, but does follow results for water and methane hydrate mixtures\cite{mirzaeifardMultiscaleModelingSimulation2019}.

Surface tension depends on the forces of attraction between particles but also the complex behaviors that happen in the small contact layers between different materials.
Kinetic hydrate promoters and inhibitors depend on accelerating and slowing the nucleation rate by affecting the bulk mass transfer of gas to the expanding hydrate structure\cite{kvammeSmallAlcoholsSurfactants2021}.
If the ability of the gas molecule to penetrate the pre-melting layer is affected by the surface tension and dipole properties, the overall nucleation rate is also affected.
Therefore, obtaining accuracy results for surface tension is critical for downstream measures of additive performance\cite{mirzaeifardCharacterizationNucleationMethane2019}.

\section{Conclusions}
\label{sec:conclusions}

To answer questions surrounding natural gas hydrate nucleation and growth, we studied the molecular phenomena at the interface between the hydrate phase and natural gas.
We glimpsed the interfacial structures and forces by studying the density distribution, molecular dipoles, and interfacial tension in this interface.
By understanding these properties, we can guide engineering applications on additives and surface coverings to promote or inhibit gas hydrate formation and growth.

We first used distance-based metrics to assess the equilibrium and sampling properties of our system, concluding the system has reached the desired state.
By computing the density profile as a function of distance to the interface, we established the presence of heterogeneous nucleation due to surface adsorption on the hydrate by gaseous molecules.
The density profile was fitted to a hyperbolic tangent to calculate the interfacial thickness, showing an expansion with temperature.
The molecular dipole was characterized normal and tangent to the interface, showing that water molecules are preferentially aligned normal to the interface, with oxygen atoms points towards the hydrate structure in the gas phase.
In the pre-melting layer, molecules are strongly aligned parallel to the interface to provide the highest probability of hydrogen bonding and participate in the hydrate structure.
This balancing of surface forces yields to increasing surface tension with temperature.
The unusual trend has been seen for surface stresses in water/methane hydrate systems and ice/air systems, showing the importance in distinguishing surface stress and energy.

The results herein report on the different facets of natural gas hydrates at the molecular level.
By studying the dipole and stresses, we can provide a better understanding of the possible methodologies to control hydrate formation kinetics to tune performance for various engineering applications.
Further studies are required to explore possible avenues for exploiting the observed dipole behavior to tune surface coverings and additives in the future. 

\section{Acknowledgements}
\label{sec:acknowledgments}
This work is supported by the Fonds de Recherche du Québec Nature et Technologies through the Bourse de Doctorat en Recherche and the Natural Sciences and Engineering Research Council of Canada through the Canada Graduate Scholarship Doctoral award.
The authors also acknowledge the Digital Research Alliance of Canada for computational resource grants, expertise, and support.

\bibliographystyle{unsrtnat}
\bibliography{main}

\end{document}